\documentclass[prl,twocolumn,twoside,aps,floatfix,longbibliography,natbib,nofootinbib,superscriptaddress]{revtex4-1}

\usepackage{amsmath, bm, amsfonts, amssymb}     
\usepackage{graphicx}                       
\usepackage{xfrac}                          
\usepackage{grffile}                        
\usepackage{color,colortbl}
\usepackage{xcolor}

\usepackage[colorlinks, citecolor=blue, urlcolor=red, linkcolor=blue]{hyperref}

\newcommand{\xhat}{\hat{\mathbf{x}}}
\newcommand{\yhat}{\hat{\mathbf{y}}}
\newcommand{\zhat}{\hat{\mathbf{z}}}

\newcommand{\tw}{t_{\rm w}}

\newcommand{\be}{\begin{equation}}
\newcommand{\ee}{\end{equation}}
\newcommand{\bea}{\begin{eqnarray}}
\newcommand{\eea}{\end{eqnarray}}

\newcommand{\friction}{F}
\newcommand{\load}{N}
\newcommand{\velocity}{U}
\newcommand{\FD}{F_{\rm D}}
\newcommand{\FS}{F_{\rm S}}
\newcommand{\muD}{\mu_{\rm D}}
\newcommand{\muS}{\mu_{\rm S}}
\newcommand{\Fy}{F_{\rm Y}}
\newcommand{\Anominal}{A}

\begin{document}

\title{A model of friction with plastic contact nudging:  Amontons-Coulomb laws, aging of static friction, and non-monotonic Stribeck curves with finite quasistatic limit.}
\author{Suzanne M. Fielding}
\affiliation{Department of Physics, Durham University, Science Laboratories, South Road, Durham DH1 3LE, UK}
\begin{abstract}
We introduce a model of friction between two contacting (stationary or co-sliding) rough surfaces, each comprising a random ensemble of polydisperse hemispherical bumps. In the simplest version of the model, the bumps experience on contact with each other only pairwise elastic repulsion and dissipative drag. These minimal ingredients are sufficient to capture a static state of jammed, interlocking contacting bumps, below a critical frictional force that is proportional to the normal load and independent of the apparent contact area, consistent with  the Amontons-Coulomb laws of friction. However, they fail to capture two widespread observations: (i) that the dynamic friction coefficient (ratio of frictional to normal force in steady sliding) is a roughly constant or slightly weakening function of the sliding velocity $U$, at low $U$, with a non-zero quasistatic limit as $U\to 0$, and (ii) that the static friction coefficient (ratio of frictional to normal force needed to initiate sliding) increases (``ages'') as a function of the time that surfaces are pressed together in stationary contact, before sliding commences. To remedy these shortcomings, we incorporate a single additional model ingredient: that contacting bumps plastically nudge one another slightly sideways, above a critical contact-contact load. With this additional insight, the model also captures observations (i) and (ii). 

\end{abstract}

\maketitle


The physics of friction governs the resistance of contacting surfaces to relative sliding. It dominates myriad systems, on lengthscales ranging from the microscopic, in micromachines and biological motors~\cite{urbakh2004nonlinear}, to the geophysical, in earthquake faults~\cite{scholz2019mechanics}. The macroscopic rheology of granular matter~\cite{behringer2018physics} and dense granular suspensions~\cite{seto2013discontinuous,wyart2014discontinuous} is determined largely by the friction acting microscopically between grains. 
From an economic viewpoint, it has been argued that about $1\%$ of gross national product could be saved by mitigating friction~\cite{jost2005tribology}. Although its scientific study dates back to Leonardo da Vinci in the 15th century, many outstanding challenges remain even today. For reviews, see~\cite{vanossi2013colloquium,baumberger2006solid,muser2001simple,urbakh2004nonlinear,persson2013sliding,popov2010contact}.

From a macroscopic viewpoint, the Amontons-Coulomb laws~\cite{bowden2001friction} state that the frictional force $F$
acting tangentially between two surfaces is directly proportional to the normal force $\load$ between them, and independent of their apparent contact area $\Anominal$. Furthermore, the frictional force $F=\FD$ in dynamic sliding  is roughly independent of the relative sliding velocity $\velocity$. The ratio of frictional to normal force,  encoded in the coefficient of friction, $\mu=\friction/\load$, is typically in the range $0.1-1$ (with lower values in superlubricity~\cite{dienwiebel2004superlubricity}). The frictional force $\FS$ needed to initiate sliding from rest  typically exceeds its dynamic counterpart in steady sliding,  $\muS>\muD$. Later corrections~\cite{rabinowicz1958intrinsic} to these phenomenological laws showed that the coefficient of sliding friction $\muD$ in fact tends to decrease (``weaken") slightly with increasing $\velocity$, at small $\velocity$; and that the coefficient of static friction $\muS$ often increases (``ages") over the time that  surfaces are pressed in stationary contact, before sliding commences. 

Theoretical approaches to friction range from models and~\cite{prandtl1928conceptual,tomlinson1929cvi}  quantum mechanical simulations~\cite{kuwahara2017friction,wolloch2018interfacial} of nanofriction, to molecular dynamics simulations~\cite{he1999adsorbed,muser2001simple}, to mesoscopic multi-contact models~\cite{persson1995theory,braun2009dynamics}, up to rate-and-state models, which coarse grain the surface rheology as a whole via a small number of {\em macroscopic} dynamical variables, and posit constitutive equations for their evolution~\cite{ruina1983slip,dieterich1979modeling}. Among these, multi-contact models are appealing in allowing access to larger time and lengthscales than molecular simulations, while avoiding the severe constitutive assumptions of rate-and-state models. Indeed, the only assumptions are {\em mesoscopic} ones of  pairwise interactions between contacting surface bumps, with {\em macroscopic} surface rheological (frictional) behaviour then emerging out of a many-body bump level simulation. This is akin to assuming pairwise interactions in a particle level simulation of a soft material, from which macroscopic bulk rheology then emerges.

Indeed, some early attempts to understand friction invoked a multi-contact picture in which the bumps of contacting surfaces interlock~\cite{dowson1979history}. Within any such description, a key puzzle is how to explain the observation that the dynamic friction coefficient approaches a finite limit in quasistatic sliding, $\lim_{U\to 0}\muD(U)\neq0$. In particular, within such simple bump models, the work done against the elastic repulsion that resists the coming together of any two bumps should be exactly recovered by that hastening their separation as they move past each other and part, giving zero average force. (Adding dissipation confers a kinetic friction, but only at finite $U$.) 

Attempts to resolve this conundrum have variously invoked internal surface instabilities, in which motion proceeds via a series of discontinuous jumps~\cite{prandtl1928conceptual,tomlinson1929cvi}; elastohydrodynamics, with a high shear rate in a thin lubrication film between contacts~\cite{zhu2015stribeck}; and slow ``soft glassy rheology" in the nanometric ``joint" between contacts~\cite{bureau2002rheological}. In essence, these approaches involve either a finite rate of internal surface deformation processes even as $U\to 0$, or a zero rate of stress relaxation, such that the friction, scaling as the ratio of deformation rate to stress relaxation rate, remains finite as $U\to 0$.

\begin{figure}[!t]
\includegraphics[width=7.0cm]{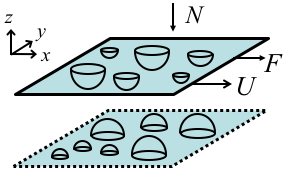}
\caption{Model of friction between opposing rough surfaces. Each surface comprises a plate studded with polydisperse hemispherical bumps. The plates are pressed together with a normal load $N$ per area, and subject to either an imposed  frictional force $F$ per area, with the sliding velocity $U$ measured; or an imposed $U$, with $F$ measured.}
\label{fig:schematic}
\end{figure}

\begin{figure}[!t]
\includegraphics[width=5.5cm]{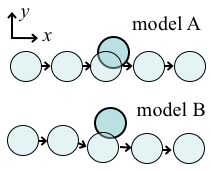}
\caption{A bump on the upper surface passes one on the lower surface. In model A, with no plastic contact nudging, the left-right symmetry results in zero average frictional force along $x$, in the quasistatic limit $U\to 0$. In  contrast, in model B, the elastic repulsion hindering contact-contact approach along $x$ exceeds that hastening separation, because the contacts meanwhile plastically nudge each other  sideways in $y$.} 
\label{fig:trajectory}
\end{figure}

In this Letter, we introduce a simple multi-contact model of friction between contacting rough surfaces, Figs.~\ref{fig:schematic} and~\ref{fig:trajectory}, and demonstrate it to capture all the key friction phenomenology discussed in paragraph 2 above. In particular, it resolves the conundrum just described via the basic insight  (made in moving from model version A to B below) that two contacting  bumps, above a critical pairwise contact-contact load, nudge each other irreversibly slightly sideways as they pass (for cosliding surfaces) or are pressed together (in stationary contacting surfaces).  In this way, for cosliding surfaces, the work done against the elastic repulsion resisting their approach (in the direction $x$ of overall sliding) exceeds that hastening their separation as they later part, because they have meanwhile nudged each other slightly sideways 
along $y$. See Fig.~\ref{fig:trajectory} (bottom). This confers a finite dynamic friction coefficient $\muD$, even in quasistatic sliding, $U\to 0$; and predicts $\muD$ to decrease slightly at small $U$, before it increases again at large $U$, as dissipation leads to greater relative dynamic surface drag. It also captures the widely observed ageing over time of the coefficient of static friction $\muS$ (with $\muS>\lim_{U\to 0}\muD$) for surfaces pressed together in stationary contact, as contacts progressively nudge sideways, allowing closer surface meshing. In contrast, model A, with no plastic contact nudging, has $\muD(U=0)=0$ and no ageing of $\muS$.

The notion of irreversible plastic contact deformation is consistent with the insight that the real contact area between rough surfaces, in being composed of many contacting microscopic asperities, is much less than the apparent contact area~\cite{bowden2001friction,weber2018molecular}.  In this way, the actual load per area borne by the contacts greatly exceeds the apparent load per area across the surfaces, and can exceed the threshold for plastic contact yielding. Indeed, the sideways plastic {\em nudging} of bumps that otherwise keep their shape, as adopted here, is intended as a simple cartoon of sideways plastic bump {\em deformation}, which would however be much more difficult to model. We do not expect this simplification to affect our conclusions: the argument summarised in Fig.~\ref{fig:trajectory} holds in either case.

{\it Model A---} considers two opposing surfaces,  each comprising a flat plate studded with polydisperse hemispherical bumps, Fig.~\ref{fig:schematic}. The bump radii $R$ are drawn from a top hat distribution of mean $R_0$ and half-width $\alpha R_0$.  The fractional bump area coverage of each surface is $\phi$.  The bump locations in the $xy$ plane of each surface are first initialised at random, so inevitably with some $xy$ overlaps within each surface. These are removed numerically via steepest descent dynamical repulsion in $xy$. Thereafter (in model A) the bumps maintain fixed positions relative to their respective plates. (We relax this assumption in  model B below.) The upper surface, which has $P$ bumps (setting our system size), can move relative to the lower surface, which is much larger and fixed in space. 

We denote the initial separation along $z$ between the two surfaces at any $x,y$ by $h(x,y)$, and define $h_0$ to be the minimum of $h$ over $x$ and $y$.
Starting from a small initial separation $h_0$  between the surfaces at time $t=0$, the upper surface is pressed onto the lower one with a normal force per unit area $-N\zhat$ for all times $t>0$.  During an initial ``waiting time" of duration $\tw$, it is held stationary in  $xy$. For all subsequent times $t>\tw$, it is subject to either a prescribed sliding velocity that rises on a short timescale $\tau$ from rest to $U\xhat$  and is held constant thereafter, with the frictional force per unit area $F\xhat$ measured in response (imposed velocity mode); or to a prescribed constant frictional force $F\xhat$, with the sliding velocity $U\xhat$ measured in response (imposed force mode). Its $y$ position and angular orientation are fixed. 

When any two opposing bumps $i$ and $j$ make contact, they experience equal and opposite pairwise  elastic repulsion with modulus $G$, and dissipation with drag coefficient $\eta$: the force on $i$ is $\vec{f}_i=-G(1-r/(R_i+R_j))\hat{r}-\eta\vec{v}_{\rm rel}$, with $\vec{r}$ the $i\to j$ centre-centre vector and $\vec{v}_{\rm rel}=\dot{\vec{r}}$ the relative velocity of $i$ and $j$. The net force on the upper plate is the sum of those acting on its constituent bumps, plus an external force comprising (per unit area) the load $-N\zhat$, a lateral force $L\yhat$ calculated to maintain the plate's constant $y$ position, plus the frictional force $F\xhat$. $F$ is either prescribed (imposed force mode), or calculated to ensure a given imposed sliding speed (imposed velocity mode). The upper plate has mass per unit area $M$ and obeys Newton's second law, evolved numerically with timestep $\Delta t$, with results converged to  $\Delta t\to 0$. 

{\it Model B---} is the same as model A, {\em except} for a single additional physical ingredient: that contacting bumps can irreversibly nudge each other sideways. Specifically, when any bump $i$ on the upper plate experiences a net contact force $\vec{f}_i$ of magnitude $F_i$ that exceeds a threshold $\Fy$, it moves plastically relative to the plate at velocity $\vec{v}_i=(F_i-\Fy)\hat{\vec{f}}_i/\zeta$, along $\hat{\vec{f}}_i$,  with drag coefficient $\zeta$. For computational efficiency we take the bumps on the lower plate to be much less compliant, so not subject to plastic nudging. We do not expect this simplification to affect any of our key physical conclusions. 

\begin{figure}[!t]
\includegraphics[width=9.0cm]{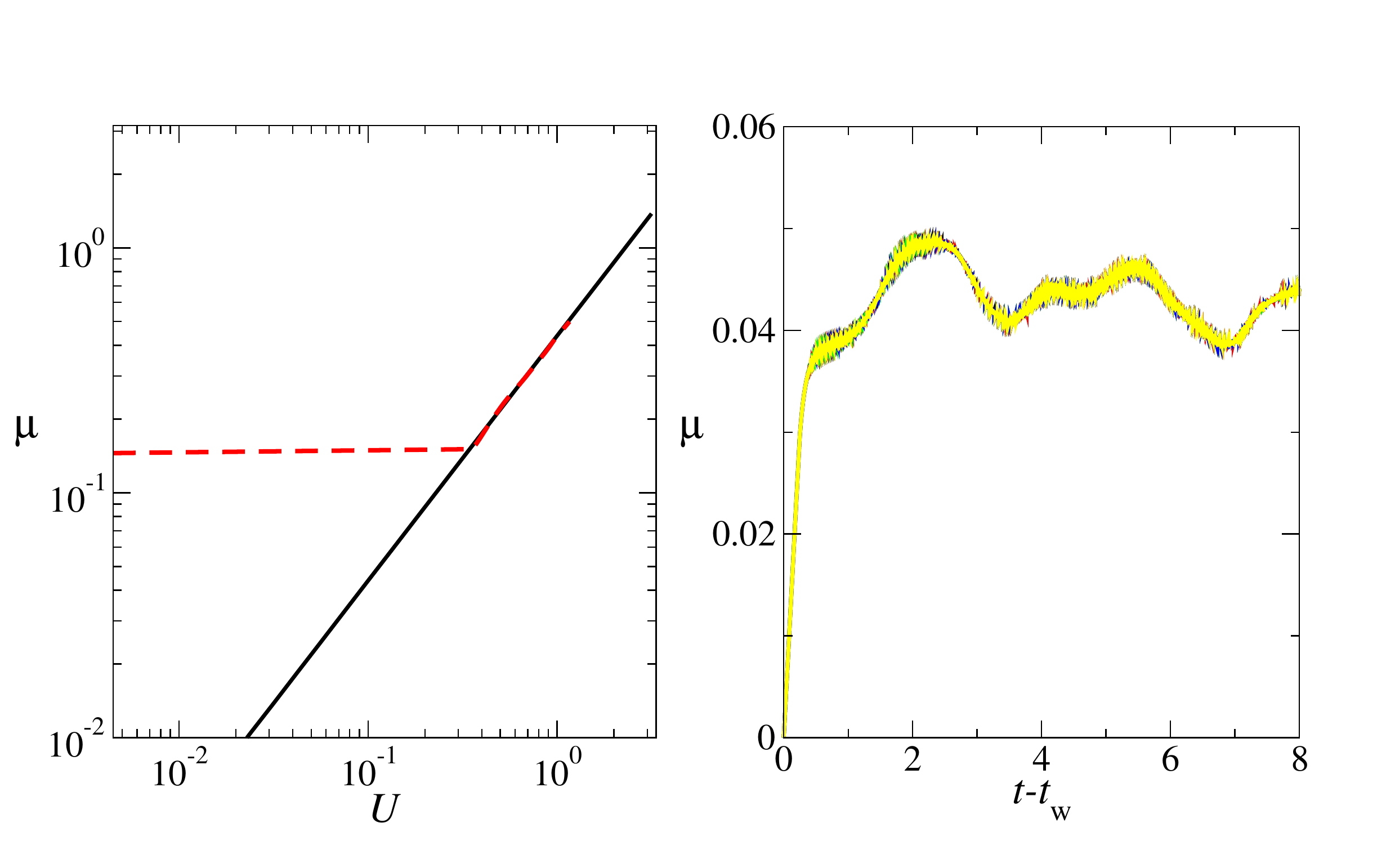}
\caption{Model A. {\bf Left)} Solid line: Stribeck curve of dynamical friction coefficient $\muD$ vs. imposed sliding velocity $U$. Dashed line: counterpart curve obtained by instead imposing the frictional force, $\tw=10.0$.  {\bf Right)} Dynamical friction coefficient vs. time $t-\tw$ since sliding starts for system size $P=4096$ at $U=0.1$, after a waiting time $\tw$ of stationary contact, with $\tw=10^0,10^{1/2},10^1,10^{3/2},10^2$  (curves indistinguishable); each curve averaged over 10 seed values}
\label{fig:Fig1}
\end{figure}

The parameters of the model are: the number of bumps $P$, average bump radius $R_0$, bump polydispersity $\alpha$, bump area coverage of the plates  $\phi$, modulus of the bump material $G$, dissipation between contacting bumps  $\eta$,  plate mass $M$ and (in model B only) the threshold force $F_{\rm y}$ and above-threshold drag $\zeta$. The parameters  of the experimental protocol are: the initial plate separation $h_0$, imposed normal load $N$, how long the plates are pressed together in contact before sliding starts $\tw$, imposed tangential velocity $U$ (or frictional force $F$), and velocity rise time $\tau$. Of these, three set our units of mass, length and time: $R_0=1$, $\eta=1$, $N=1$ (after rescaling the kinetic coefficients $\eta\to\eta\sqrt{G}$ and $\zeta\to\zeta\sqrt{G}$ for convenience).  Our results are independent of $M,h_0,\tau, P$ and $G$, in the physically relevant limit of large $P$ and $G$. Unless otherwise stated we  set $M=1.0$, $h_0=0.5$,   $\tau=1.0$, $P=512$ and $G=10^3$, $\phi=0.3$ and $\alpha=0.5$.  The remaining parameters to explore are then (of the protocol) $U$ (or $F$) and $\tw$; and (in model B)  $\Fy$ and $\zeta$.

{\it Results ---} We start by describing the predictions of model A in imposed velocity mode. Fig.~\ref{fig:Fig1}, left shows as a solid line the dynamical friction coefficient $\muD$ {\it vs.}  the imposed adimensional sliding velocity $U$, averaged over many time units once a state of statistically steady sliding is attained. In the lubrication literature, this is called the Stribeck curve. In model A, $\muD$ is linear in $U$, with zero dynamical friction in the limit of quasistatic sliding, $U\to 0$. The reason is as follows. Consider any two opposing bumps as they approach, pass, then move apart, along the direction $x$ of plate sliding. In model A, these experience on contact only elastic repulsion and (at finite $U$) dissipation. The resulting left-right symmetry, Fig.~\ref{fig:trajectory} (top), means that the elastic force hindering their approach is exactly canceled by that hastening their separation, giving zero friction on average for $U\to 0$. 

\begin{figure}[!t]
\includegraphics[width=9.0cm]{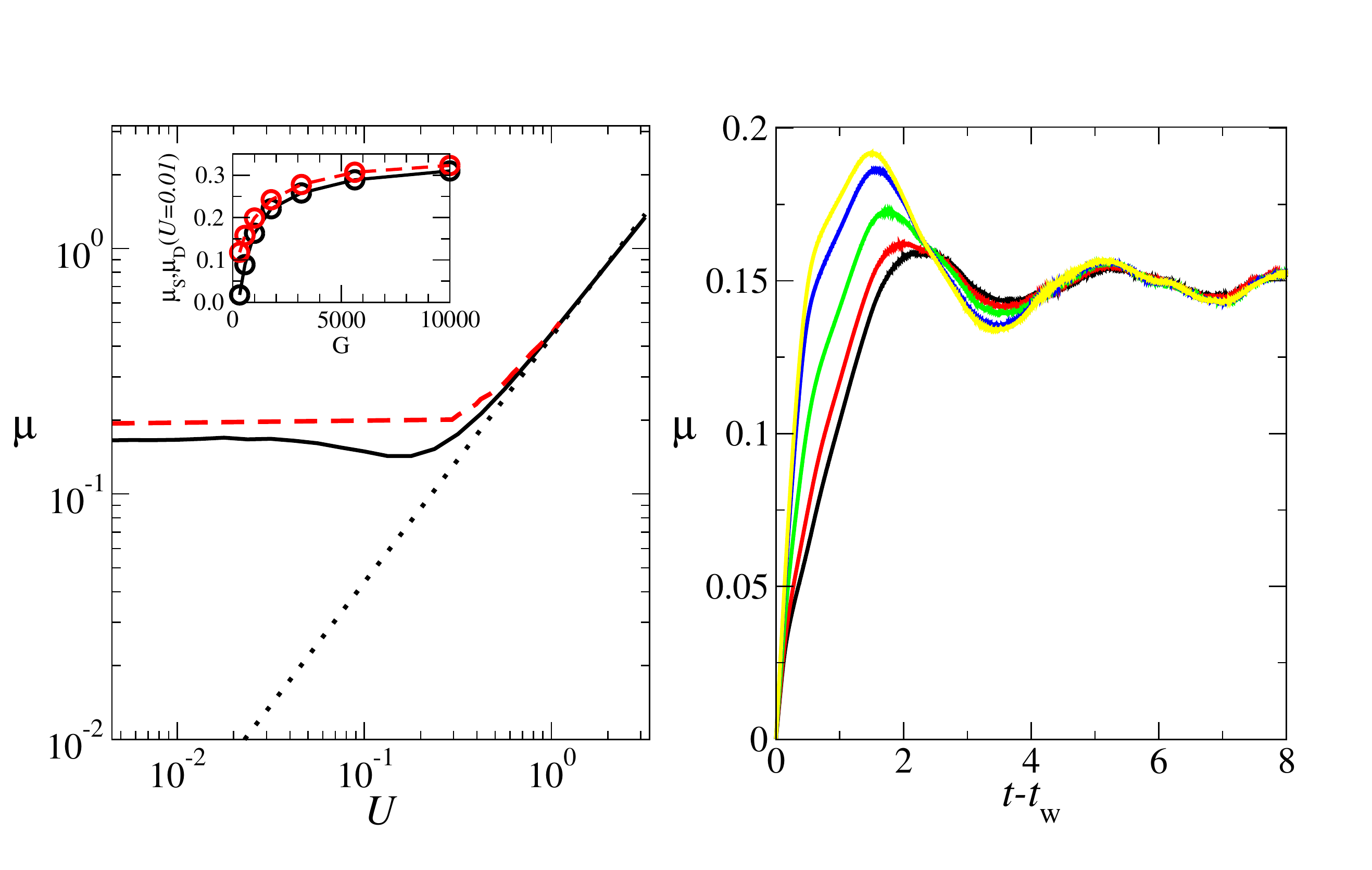}
\caption{Model B with $\zeta=10.0,\Fy=40.0$. {\bf Left)} Solid line: Stribeck curve. (Dotted: Stribeck curve of model A for comparison.) Dashed: counterpart curve when imposing the frictional force, $\tw=10.0$. Inset: value of $\muS$ (dashed) and of $\muD$ nearing the quasistatic limit, $U=0.01$ (solid) vs modulus $G$.   {\bf Right)} Dynamical friction coefficient vs. time $t-\tw$ since sliding starts for system size $P=4096$ at $U=0.1$, after a waiting time $\tw$ of stationary contact, with $\tw=10^0,10^{1/2},10^1,10^{3/2},10^2$ in order of increasing maximum; each curve averaged over 10 seed values. }
\label{fig:Fig2}
\end{figure}

The dynamical friction coefficient as a function of time $t-\tw$ since the imposition of sliding is shown in Fig.~\ref{fig:Fig1}, right. This displays a slight overshoot before settling onto the final state of statistically steady sliding. The overshoot height is independent of the waiting time $\tw$ before sliding starts: model A lacks frictional ageing. Each curve $\mu(t-\tw)$ is averaged over ten runs with different bump initialisations. Averaging over more runs would further smooth the irregular nature of the signal.

If we instead impose the frictional force $F$, we find a critical threshold force $\FS$ below which there is a jammed state of interlocking bumps with no relative sliding. Above this threshold, a state of statistically steady sliding is attained, with the same dynamical friction coefficient as in the mode of imposed velocity $U$ explored above. See Fig.~\ref{fig:Fig1}, left, red dashed line.

Compared with the key friction phenomenology summarised in paragraph 2 above, model A has some significant shortcomings. First, it fails to capture the widespread observation that $\muD$ is relatively independent of $U$ in slow sliding, and with a finite value in the limit of quastistatic sliding, $U\to 0$. Second, it fails to predict the widely observed slight weakening of $\muD$ with increasing $U$. Third, it fails to predict the ageing of $\muS$ with increasing time $\tw$ that surfaces are pressed together in stationary contact before sliding commences. To remedy these shortcomings, we turn now to model B.

\begin{figure}[!t]
\includegraphics[width=9.0cm]{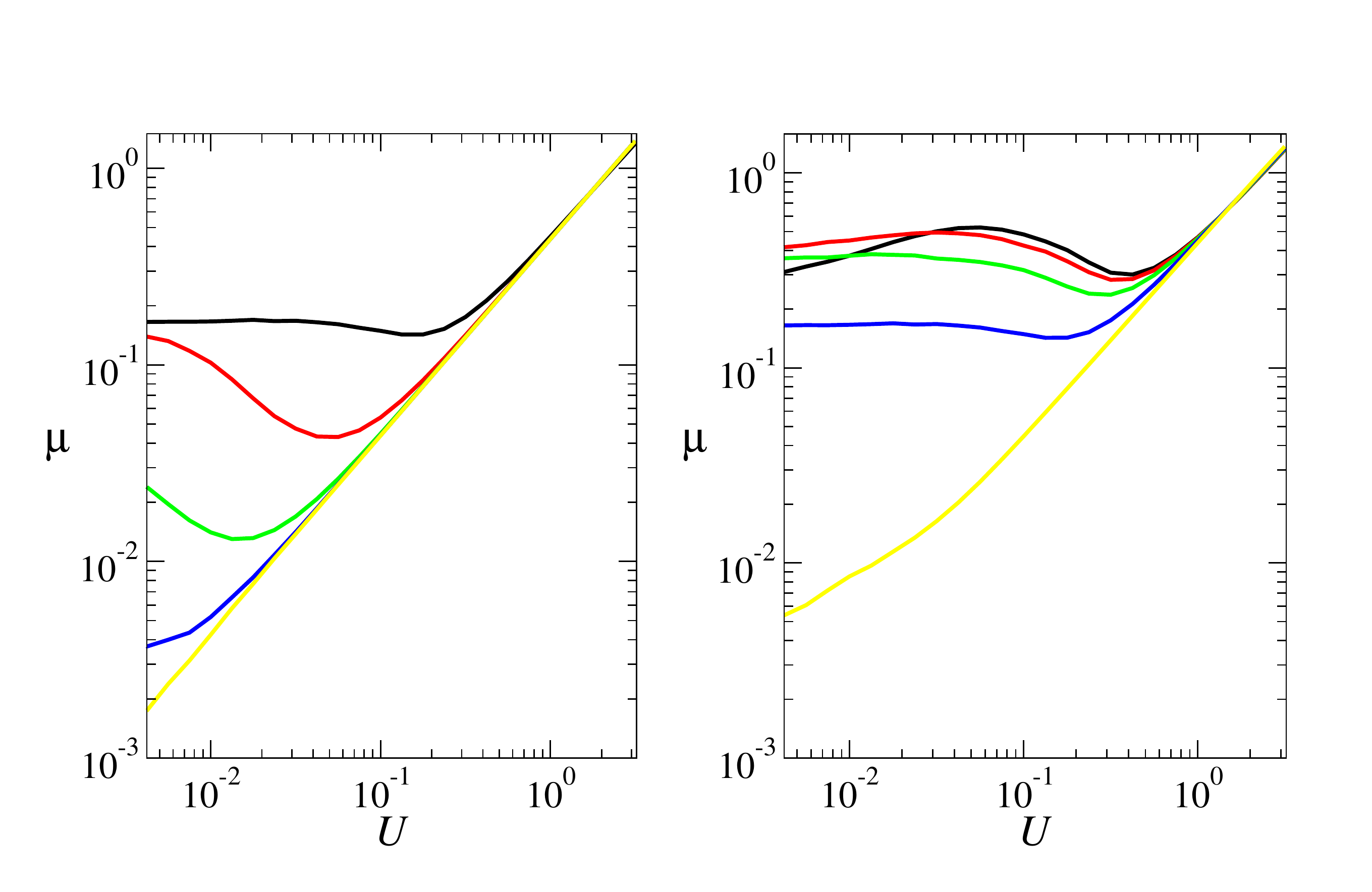}
\caption{Model B: Stribeck curves of dynamical friction coefficient $\mu_{\rm D}$ vs. imposed sliding velocity $U$. {\bf Left)} For fixed threshold $F_{\rm Y}=40.0$ and varying drag coefficient $\zeta=10^n$ for plastic contact nudging, with $n=1,2,3,4,100$ in curves downwards. For $n=100$, nudging is suppressed, and model B reduces to model A. {\bf Right)} For fixed $\zeta=10.0$ and varying $\Fy=5.0,10.0,20.0,40.0,80.0$ in curves downward at $U=0.1$.}
\label{fig:Fig4}
\end{figure}

Model B's Stribeck curve of dynamical friction {\em vs.} imposed velocity is shown by the black solid line in Fig.~\ref{fig:Fig2}, left. (The counterpart curve for model A is shown by the dotted line for comparison.) In important contrast to model A, model B now predicts $\muD$ to be relatively independent of $U$ in slow sliding, with a finite quastistatic limit as $U\to 0$.  The reason is as follows. As two contacts pass each other along $x$, they also nudge each other slightly sideways along $y$. The elastic force resisting their approach along $x$ thereby exceeds that hastening their later coming apart, because they have meanwhile nudged each other sideways. Model B also successfully predicts a slight weakening of $\muD$ with increasing $U$, before $\muD$ increases again at high $U$ due to dissipation. In imposed force mode, a static state of stationary interlocking bumps is predicted below a threshold $\muS$ that slightly exceeds $\muD(U=0)$ (red dashed line in Fig.~\ref{fig:Fig2}, left).

Model B's dynamical friction coefficient as a function of time $t-\tw$ since sliding commences is shown in Fig.~\ref{fig:Fig2}, right. As in model A, this display a slight overshoot before settling onto the final state of statistically steady sliding. In notable contrast to model A, however, the height of the overshoot increases with the waiting time $\tw$ before sliding starts: model B captures frictional ageing via contacts progressively nudging each other slightly sideways as surfaces are pressed together in stationary contact, allowing gradually closer surface meshing.

In Fig.~\ref{fig:Fig4}, we explore the dependence of model B's Stribeck curves on the threshold $\Fy$ and drag coefficient $\zeta$ for contact nudging. Increasing the drag coefficient $\zeta$ simply makes nudging more  sluggish and shifts its effects to lower $U$ (left panel). Indeed, taking $\zeta\to \infty$ at fixed $U$, the contacts are unable to nudge at all, and model B reverts to model A. Fig.~\ref{fig:Fig4}, right shows the effect of changing the nudging threshold $\Fy$. As $\Fy\to \infty$, contacts are again unable nudge, recovering model A. For the smaller values of $\Fy$  explored, $\mu$ is relatively independent of $\Fy$. (For very small $\Fy$, in the limit $U\to 0$, we might again expect a return to zero $\muD$, because zero force is needed to nudge the contacts. This regime is likely to be unphysical, however: we do not expect contacts to nudge plastically at arbitrarily small forcing.)

We have verified the friction coefficient to be independent of system size by performing runs for $P=1024$ (data not shown) as well as the default $P=512$. In this way, our model  predicts the frictional force to be independent of the apparent area of contact between surfaces, consistent with the Amontons-Coulomb laws.

In quastistatic sliding, the only parameters with dimensions of force that can determine the frictional force $F$ in model B are the normal load $N$, particle modulus $G$, and nudging threshold $\Fy$. We have just shown the coefficient of friction to be relatively independent of $\Fy$, for moderate $\Fy$. In the inset of Fig.~\ref{fig:Fig2}, left, we further demonstrate the coefficient of friction to approach a value independent of the contact modulus $G$ at large $G\approx 10^4$. ($G=10^4$ is however very costly numerically and our default value otherwise is $G=10^3$.) In this way, our model predicts the frictional force to be proportional to the load $N$, consistent with the Amonton-Coulomb laws.

{\it Conclusions ---} We have introduced a multi-contact model of rough contacting surfaces, and showed it to capture the key phenomenology of static and dynamic friction. Our choice of hemispherical bumps with a top-hat polydispersity distribution of radii, randomly placed on the plates, necessarily corresponds to a particular statistics of surface roughness. In practice, rough surfaces are often self-
similar in nature, and it would be interesting to consider a correlated placing of the bumps with a fractal surface morphology.
In practice, such surface statistics~\cite{pastewka2013finite} are likely to determine the detailed functional form of the weakening dynamical friction coefficient $\muD(U)$ and ageing static friction coefficient $\muS(\tw)$: a study we defer to future work.  Our treatment does not capture the ageing sometimes observed of the {\em dynamic} friction coefficient $\muD$, instead admitting a steady Stribeck curve $\muD(U)$: the ensemble of bumps attains a statistically steady state at any $U$, even though any individual bump is nudged around on contact with others. Ageing of $\muD$ could potentially be incorporated via a plastic {\em wearing downs} of bump on contact. It would also be interesting to investigate the effects of bump-bump adhesion on ageing~\cite{tian2020linear} in this model.

Our study captures much of the observed phenomenology of both dry solid surfaces, with dissipation due to plasticity within contacts, as well as lubricated ones, with dissipation in fluid films. However, it does not consider in detail the hydrodynamic interactions between the contacts, which should properly be modelled via Stokes flow in the lubrication regime~\cite{zhu2015stribeck,hamrock2004fundamentals,warren2016sliding}. Instead, we have adopted a simplified drag term $-\eta\vec{v}_{\rm rel}$ between close contacts. Both approaches will, however, give Stribeck curves with increasing $\mu(U)$ at high $U$. Our key focus has instead been on small to moderate $U$.

Throughout, we have focused on materials with contacts sufficiently far apart that elastic stress propagation between them, within each surface, can be neglected. Including this could address the initial onset of sliding via spatio-temporally propagating rupture fronts~\cite{svetlizky2019brittle,xia2004laboratory,rubinstein2004detachment,ben2010dynamics,passelegue2013sub,de2019collective}.

It is worth noting finally a similarity between (a) the tribology of opposing surfaces comprising disordered ensembles of soft bumps and (b) the rheology of yield stress fluids comprising disordered packings of soft particles~\cite{bonn2017yield}. The latter are often termed `{\em elastoviscoplastic}', with analogy to the {\em elastic} contact forces, {\em viscous} drag forces, and {\em plastic} contact nudging here. The force-velocity  ``Stribeck'' curves of (a) and the stress/strain-rate ``flow curves'' of (b) both show jamming below a critical forcing, and sliding/shearing at larger forcing. The frictional force in (a) and shear stress in (b) show an overshoot vs. the time $t-\tw$ since in the inception of sliding/shearing, with a height that grows with increasing waiting time $\tw$ before sliding/shear commenced. Our model of (a) also predicts creep-and-yielding curves following the imposition of a step frictional force  (not shown) with the same rich qualitative features as shear rate vs time in (b) following the imposition of a step shear stress, the study of which we defer to future work.

{\it Acknowledgements ---} The author thanks Mike Cates and Patrick Warren for helpful comments on the manuscript. This project has received funding from the European Research Council (ERC) under the European Union's Horizon 2020 research and innovation programme (grant agreement No. 885146).


%

\end{document}